\documentclass[useAMS,usenatbib]{mn2e}
\usepackage {graphicx}

\title[Thermal Stability of Thin Accretion Discs]
{Effects of Stress Evolution Process on the Thermal Stability
of Thin Accretion Discs}

\author[Da-Bin Lin, Wei-Min Gu, and Ju-Fu Lu]
{Da-Bin Lin, Wei-Min Gu\thanks{E-mail:
guwm@xmu.edu.cn}, and Ju-Fu Lu\\
Department of Physics and Institute of Theoretical Physics and
Astrophysics, Xiamen University, Xiamen, Fujian 361005, China}
\begin{document}

\date{}

\maketitle

\begin{abstract}
The stress evolution process is taken into account in the linear
stability analysis of standard thin accretion discs.
We find that the growth rate of thermally unstable modes can
decrease significantly owing to the stress delay,
which may help to understand the quasi-periodic variability of GRS 1915+105.
We also discuss possible application of stress evolution
to the stability of Shapiro-Lightman-Eardley disc.
\end{abstract}


\section{Introduction}

The standard thin accretion disc (\citealp{Shakura73})
has been extensively applied to a variety
of astrophysical systems, such as active galactic nuclei,
X-ray binaries, and cataclysmic variable stars.
The theory of the standard thin disc, which is based on the $\alpha$
stress prescription, predicts that the disc will
suffer both thermal and viscous instability when the pressure
is dominated by radiation pressure.
However, observations of X-ray binaries show little evidence
for such instabilities except for GRS 1915+105 (hereafter GRS 1915).
Since the stability is related to the description of stress,
the standard $\alpha$ stress prescription
was modified in some previous works to assure the disc to be stable
no matter whether gas or radiation pressure dominates.
For instance, \citet{Lightman74} introduced a so-called $\beta$
stress prescription, i.e., the stress is proportional to the gas pressure
rather than the sum of the gas and radiation pressure.
With the $\beta$ stress, the disc will remain stable
even for the case that radiation pressure dominates over gas pressure.
Furthermore, some other stress prescriptions were introduced by assuming
that the stress is relevant to both the total pressure and the gas pressure
(e.g., \citealp{Kato08}, p.~108).
As pointed out by \citet{Gierli04}, however,
the $\beta$ stress would be in conflict with observations as follows.
Observations show that the colour
temperature correction remains stable for a wide range of luminosity,
whereas discs with $\beta$ stress should have a colour temperature correction
that changes from $\sim 1.8$ to $\sim 2.7$ as the accretion rate increases.
Conversely, the standard $\alpha$ prescription may predict that
the correction should remain stable at $\sim 1.8$ for varying accretion rates.
Moreover, some recent simulations (e.g., \citealp{Hirose09})
showed that the stress is roughly proportional to the sum of gas and
radiation pressure. Thus, the standard $\alpha$ prescription seems to be
more acceptable than other prescriptions such as the $\beta$ stress.

The slim disc model was introduced by \citet{Abramowicz88},
in which it was predicted that
the thermal instability of the radiation-pressure-dominated thin
disc will trigger a limit cycle behaviour between the thin disc
and the slim disc. Such a limit cycle was confirmed by time-dependent
numerical calculations
(e.g., Szuszkiewicz \& Miller 2001; \citealp{Li07}).
On the other hand, \cite{Belloni97} 
applied the limit cycle theory to interpret the observational 
quasi-periodic variability of GRS 1915.
Recently, such limit cycle has also been applied to young radio
galaxies (e.g., \citealp{Czerny09,Wu09}).
Comparing the numerical results with the observational ones,
however, we can find the following inconsistency
between the theory and the observation. The numerical results of
Li et al.~(2007) showed that the duration of the outburst phase
$t_{\rm high}$ is less than 5 percent of the duration of the quiescent phase
$t_{\rm low}$, and the luminosity of the outburst phase $L_{\rm high}$
is around two orders of magnitude larger than that of the quiescent phase
$L_{\rm low}$; whereas observations of GRS 1915 showed that,
$t_{\rm high}$ is comparable to $t_{\rm low}$ and
$L_{\rm high}$ is only 3-20 times larger than $L_{\rm low}$
(e.g., \citealp{Belloni97, Wu10}).
Some efforts have been made to improve the theory in order to explain
observations either by some artifical viscosity prescription
(e.g., \citealp{Nayakshin00}) or by additional assumption of the
energy exchange between the disc and corona (Janiuk et al.~2002).

In the present paper, we will investigate this issue through
another way by taking the stress evolution process into account.
Our paper is organized as follows. In Section 2, we describe the
basic assumption for the stress evolution process. The time-dependent
equations and linear stability analysis are presented in Sections 3 and 4,
respectively. Conclusions and discussion are made in Section 5.

\section[]{Model}

The stress in MHD accretion discs can be expressed as
(e.g., Balbus \& Papaloizou 1999)
\begin{equation}
\tau_{r\varphi} =
<{\rho}{\delta\upsilon_r}{\delta\upsilon_{\varphi}}
- \frac{{{\delta}B_r}{{\delta}B_{\varphi}}}{4\pi}>_\varphi \ ,
\end{equation}
where ${\delta\upsilon_r}$(${{\delta}B_r}$) and
${\delta\upsilon_{\varphi}}$(${\delta}B_{\varphi}$)
are the fluctuating components of the velocity (the magnetic field)
in $r$ and $\varphi$ direction, respectively.
The angled bracket notation denotes the azimuthal average.

The above formula is a general form and is complicated for calculation.
In the study of accretion discs,
a more practical formula, i.e., the $\alpha$ stress prescription (\citealp{Shakura73}),
is widely adopted:
\begin{equation}
\tau_{r\varphi}=-{\alpha}p_{\rm tot} \ ,
\end{equation}
where $p_{\rm tot}$ is the total pressure, and $\alpha$ is regarded
as a constant less than unity.
With the above simple form, the $\alpha$ stress prescription has been
extensively applied to the research of accretion discs,
even in the disc stability analysis. But in the real system,
when the pressure varies, it will take a certain period
for the stress to adjust.
As mentioned in \S~1, the stability is sensitive to the description
of stress, thus the stress evolution process
may have essential effects on the stability.
In this paper, we will modify the standard $\alpha$ stress prescription
by taking the stress delay into account.

Generally, the growth rate of the stress should be related to 
$\tau_{r\varphi}^{\rm exp}$ and $\tau_{r\varphi}$,
where $\tau_{r\varphi}^{\rm exp}$ is the expected 
stress and $\tau_{r\varphi}$ is the current stress.
Thus we have the following general function for the evolution of stress:
\begin{equation}
\frac{{\partial}\tau_{r\varphi}}{{\partial}t}+\upsilon_r\frac{{\partial}
\tau_{r\varphi}}{{\partial}r}=f(\tau_{r\varphi}^{\rm exp}, \tau_{r\varphi})
\ ,
\end{equation}
where $\upsilon_r$ is the radial velocity.
In a quasi-stationary state, the function $f$ should match
the following conditions:
\begin{equation}
\left \{
\begin{array}{cc}
f(\tau_{r\varphi}^{\rm exp}, \tau_{r\varphi}) < 0;
& \tau_{r\varphi} > \tau_{r\varphi}^{\rm exp} \\
f(\tau_{r\varphi}^{\rm exp}, \tau_{r\varphi}) = 0;
& \ \ \tau_{r\varphi} = \tau_{r\varphi}^{\rm exp} \ . \\
f(\tau_{r\varphi}^{\rm exp}, \tau_{r\varphi}) > 0 ;
& \tau_{r\varphi} < \tau_{r\varphi}^{\rm exp}
\end{array} \right.
\end{equation}
In the present paper, we adopt a simple form as follows:
\begin{equation}
f(\tau_{r\varphi}^{\rm exp}, \tau_{r\varphi}) =
\frac{\tau_{r\varphi}^{\rm exp}-\tau_{r\varphi}}{t_{\rm {ps}}} \ ,
\end{equation}
where $t_{\rm ps}$ is the timescale of stress delay for varying pressure.
We further adopt $\tau_{r\varphi}^{\rm exp} = -\alpha{p_{\rm tot}}$
according to the $\alpha$ prescription, where $\alpha$ is a constant.
Combining equations (3) and (5),
the evolution equation of the stress tensor is
\begin{equation}
\frac{{\partial}\tau_{r\varphi}}{{\partial}t}
+ \upsilon_r\frac{{\partial}\tau_{r\varphi}}{{\partial}r}
= \frac{-{\alpha}p_{\rm {tot}}
- \tau_{r\varphi}}{t_{\rm {ps}}} \ .
\end{equation}

\section{Equations}

We first present the time-dependent equations of accretion discs
for stability analysis.
We consider an axisymmetric, Keplerian rotating disc
($\Omega=\Omega_{\rm K}$) under cylindrical coordinates
($r$, $\varphi$, $z$).
The basic vertically integrated equations are described as follows
(e.g., Kato et al.~2008):
\begin{equation}\label{MassE}
\frac{{\partial}\Sigma}{{\partial}t}
+ \frac{1}{r}\frac{{\partial}}{{\partial}r}(r{\Sigma}\upsilon_r) = 0 \ ,
\end{equation}
\begin{equation}\label{AzimuthalE}
{\Sigma}\upsilon_r\frac{1}{r}\frac{\partial}{{\partial}r}(r^2\Omega)
= \frac{1}{r^2}\frac{\partial}{{\partial}r}(r^2T_{r\varphi}) \ ,
\end{equation}
\begin{equation}\label{VerticalE}
\Omega_{\rm K}^2H^2 = \frac{\Pi}{\Sigma} \ ,
\end{equation}
\begin{equation}\label{EnergyE}
\frac{{\partial}E}{{\partial}t}
- (E+\Pi)\frac{{\partial}\ln\Sigma}{{\partial}t}
+ \Pi\frac{{\partial}{\ln}H}{{\partial}t}
= Q_{\rm {vis}}^+ - Q_{\rm {adv}}^- - Q_{\rm {rad}}^- \ ,
\end{equation}
\begin{equation}\label{StateE}
\Pi = \Pi_{\rm {gas}}+\Pi_{\rm {rad}} = \frac{k_{\rm B}}
{\mu m_{\rm H}}{\Sigma}T + \frac{2a}{3}T^4H \ ,
\end{equation}
where $\Sigma = 2 \rho H$ is the surface mass density,
$\Pi= 2 p_{\rm tot} H$, $T_{r\varphi} = 2\tau_{r\varphi} H$, and $E$
are the vertically integrated pressure, stress, and internal energy,
respectively. $T$ is the temperature, $H$ is the half-thickness
of the disc, $k_{\rm B}$ is Boltzmann constant, $\mu$ is the mean
molecular weight, $m_{\rm H}$ is the hydrogen mass, and $a$ is
radiation constant.
The quantities
$Q_{\rm vis}^+$, $Q_{\rm adv}^-$, and $Q_{\rm rad}^-$ represent the viscous
heating, advective cooling, and radiative cooling rates, respectively.
The expressions of $E$, $Q_{\rm vis}^+$, $Q_{\rm adv}^-$,
and $Q_{\rm rad}^-$ are the following:
\begin{displaymath}
E= \left [ 3(1-\beta)+\frac{\beta}{\gamma-1} \right ] \Pi \ ,
\end{displaymath}
\begin{displaymath}
Q_{\rm {vis}}^+ = T_{r\varphi}(r\frac{d\Omega}{dr}) \ ,
\end{displaymath}
\begin{displaymath}
Q_{\rm {adv}}^- = \upsilon_r\left [
-(E+\Pi)\frac{\partial{\ln\Sigma}}{{\partial}r}
+ \frac{\partial{E}}{{\partial}r}
+ \Pi\frac{\partial{\ln{H}}}{{\partial}r}\right ] \ ,
\end{displaymath}
\begin{displaymath}
Q_{\rm {rad}}^- = \frac{16acT^4}{3\bar{\kappa}{\Sigma}} \ ,
\end{displaymath}
where $\beta$ is defined as the ratio of the gas pressure
to the total pressure, i.e., $\beta \equiv \Pi_{\rm gas}/\Pi$,
$\gamma$ is the ratio of specific heating, and $\bar{\kappa}$ is
the opacity.

Since the timescale for establishment of vertical equilibrium is
the dynamic timescale, which is much shorter than the thermal timescale,
we modify the stress evolution process (Eq.~[6]) as
\begin{equation}\label{StressE}
\frac{{\partial}T_{r\varphi}}{{\partial}t}
+ \upsilon_r\frac{{\partial}T_{r\varphi}}{{\partial}r}
= \frac{-\alpha\Pi-T_{r\varphi}}{t_{\rm {ps}}} \ .
\end{equation}

\section{Stability analysis}

Now we have the set of six equations, i.e., Eqs.~(7)-(12) for linear
stability analysis, and the corresponding six physical quantities
are $\Sigma$, $\Pi$, $\upsilon_r$, $H$, $T$, and $T_{r\varphi}$.  
With subscript 0 denoting the quasi-stationary solution, and subscript 1
denoting perturbation quantities, we define the following dimensionless
variables:
\begin{displaymath}
\sigma = \frac{\Sigma_1}{\Sigma_0},\
\xi = \frac{\Pi_1}{\Pi_0},\
u = \frac{\upsilon_{r, 1}}{r\Omega},\
h = \frac{H_1}{H_0},\
m = \frac{T_{1}}{T_0},\
l = \frac{T_{r\varphi, 1}}{-\alpha\Pi_0}.
\end{displaymath}
With the assumption that all the perturbation quantities are proportional
to $\exp(\omega t - ikr)$, 
the perturbed equations corresponding
to Eqs.~(7)-(12) can be written as follows:
\begin{equation}\label{MassP}
\frac{\omega}{\Omega}\sigma-ikru=0 \ ,
\end{equation}
\begin{equation}\label{AzimuthalP}
\frac{\kappa^2}{2\Omega^2}u
= i\alpha kr(\frac{1}{r{\Omega}})^2\frac{\Pi_0}{\Sigma_0}l \ ,
\end{equation}
\begin{equation}\label{VerticalP}
2h=\xi-\sigma \ ,
\end{equation}
\begin{eqnarray}\label{EnergyP}
\frac{\omega}{\Omega}[3(1-\beta)+\frac{\beta}{\gamma-1}]\xi
+ \frac{\omega}{\Omega}{\beta}\frac{4-3\gamma}{\gamma-1}
\frac{1-\beta}{4-3\beta}(-3\xi+4\sigma-h)
\nonumber \\
- \frac{\omega}{\Omega}(4-3\beta+\frac{\beta}{\gamma-1})\sigma
+ \frac{\omega}{\Omega}h
= -\frac{3}{2}{\alpha}l - 6\alpha m + \frac{3}{2}\alpha \sigma \ ,
\end{eqnarray}
\begin{equation}\label{StateP}
\xi = \beta(\sigma+m)+(1-\beta)(4m+h) \ ,
\end{equation}
\begin{equation}\label{StressPS}
l = \frac{1}{{\omega}t_{\rm {ps}}+1}\xi
-({\Omega}t_{\rm {ps}})\frac{{\partial}{\ln}\Pi_0}
{{\partial}{\ln}r}\frac{1}{{\omega}t_{\rm {ps}}+1}u \ ,
\end{equation}
where ${\kappa}$ is the epicyclic frequency defined as
${\kappa^2} \equiv 2\Omega(2\Omega+r{d\Omega}/{dr})$.

Before numerical calculations, we would point out that,
for long-wavelength unstable modes,
the second term on the right hand side of Eq.~(18) is significantly
less than the first term, thus the equation can be reduced to
\begin{equation}
l = \frac{1}{{\omega}t_{\rm {ps}}+1}\xi \ .
\label{StressP}
\end{equation}
The reason for the above simplification is as follows.
Substituting equation (\ref{AzimuthalP}) into (\ref{StressPS}),
we will have
\begin{equation}\label{Reduction}
\left [\frac{\kappa^2}{2\Omega^2}\frac{({r{\Omega}})^2{\Sigma_0}}
{i\alpha kr{\Pi_0}}+({\Omega}t_{\rm {ps}})\frac{{\partial}{\ln}\Pi_0}
{{\partial}{\ln}r}\frac{1}{{\omega}t_{\rm {ps}}+1} \right ]
u
=  \frac{1}{{\omega}t_{\rm {ps}}+1}\xi,
\end{equation}
For standard thin discs,
${{\partial}{\ln}\Pi_0}/{{\partial}{\ln}r}$
and ${\kappa^2}/{(2\Omega^2)}$ are around unity.
Thus, for long-wavelength unstable modes, on the left hand side of Eq.~(20),
the ratio of the second term to the first term is
\begin{eqnarray}
\left|
\left[ ({\Omega}t_{\rm {ps}})\frac{{\partial}{\ln}\Pi_0}
{{\partial}{\ln}r}\frac{1}{{\omega}t_{\rm {ps}}+1} \right]/
\left [ \frac{\kappa^2}{2\Omega^2}\frac{({r{\Omega}})^2{\Sigma_0}}
{i\alpha kr{\Pi_0}} \right ] \right| \nonumber \\
\approx
\left| \frac{{\partial}{\ln}\Pi_0}{{\partial}{\ln}r}\right|
\left| \frac{1}{{\omega}t_{\rm {ps}}+1}\frac{1}{\kappa^2
/({2\Omega^2})}\right|
\left| 2\frac{t_{\rm {ps}}}{t_{\rm {th}}}\frac{H}{r} \right|
\left| kH \right|
{\ll}1 \ .
\end{eqnarray}
Thus, equation (\ref{Reduction}) can be simplified as
\begin{displaymath}
\left [ \frac{\kappa^2}{2\Omega^2}\frac{({r{\Omega}})^2{\Sigma_0}}
{i\alpha kr{\Pi_0}} \right ]
u
= \frac{1}{{\omega}t_{\rm {ps}}+1}\xi \ ,
\end{displaymath}
which is equivalent to Eq.~(19).

For non-trivial solutions of $\sigma$, $\xi$, $u$, $h$, $m$, and $l$,
we obtain the following dispersion relation from equations
(\ref{MassP})-(\ref{StateP}) and (\ref{StressP}),
\begin{eqnarray}\label{DissipationE}
c_1(\frac{\omega}{\Omega})^3
+ c_2(\frac{\omega}{\Omega})^2
+ c_3(\frac{\omega}{\Omega})
+ c_4 = 0 \ ,
\end{eqnarray}
where
\begin{displaymath}
c_1 = ({\Omega}t_{\rm {ps}})\left ( \frac{7-6\beta}{2}+\frac{\beta}{\gamma-1}
- \frac{7}{2}\beta\frac{4-3\gamma}{\gamma-1}\frac{1-\beta}{4-3\beta} \right )
\ ,
\end{displaymath}
\begin{eqnarray*}
c_2= \left ( \frac{7-6\beta}{2}+\frac{\beta}{\gamma-1}
- \frac{7}{2}\beta\frac{4-3\gamma}{\gamma-1}\frac{1-\beta}{4-3\beta} \right )\\
+ \frac{3\alpha(1+\beta)}{(4-3\beta)}({\Omega}t_{\rm {ps}})\ ,
\end{eqnarray*}
\begin{eqnarray*}
c_3 =2\alpha (kH)^2\left( \frac{\Omega}{{\kappa}} \right)^2\left (
\frac{9-6\beta}{2}+\frac{\beta}{\gamma-1}-
\right.\\
\left.\frac{9}{2}\beta\frac{4-3\gamma}{\gamma-1}\frac{1-\beta}{4-3\beta}  \right)
- \frac{3\alpha}{2}+\frac{3\alpha_(1+\beta)}{(4-3\beta)}
\ ,
\end{eqnarray*}
\begin{displaymath}
c_4 = 3\alpha^2 (kH)^2\left( \frac{\Omega}{{\kappa}} \right)^2
\frac{2+3\beta}{4-3\beta}
\ .
\end{displaymath}
When $t_{\rm ps} \to 0$, the above dispersion relation is reduced to
the form in the $\alpha$ stress scenario (e.g., Kato et al.~2008, p.~180).

\begin{figure}
\begin{center}
{\includegraphics[width=3.00in,height=2.70in]{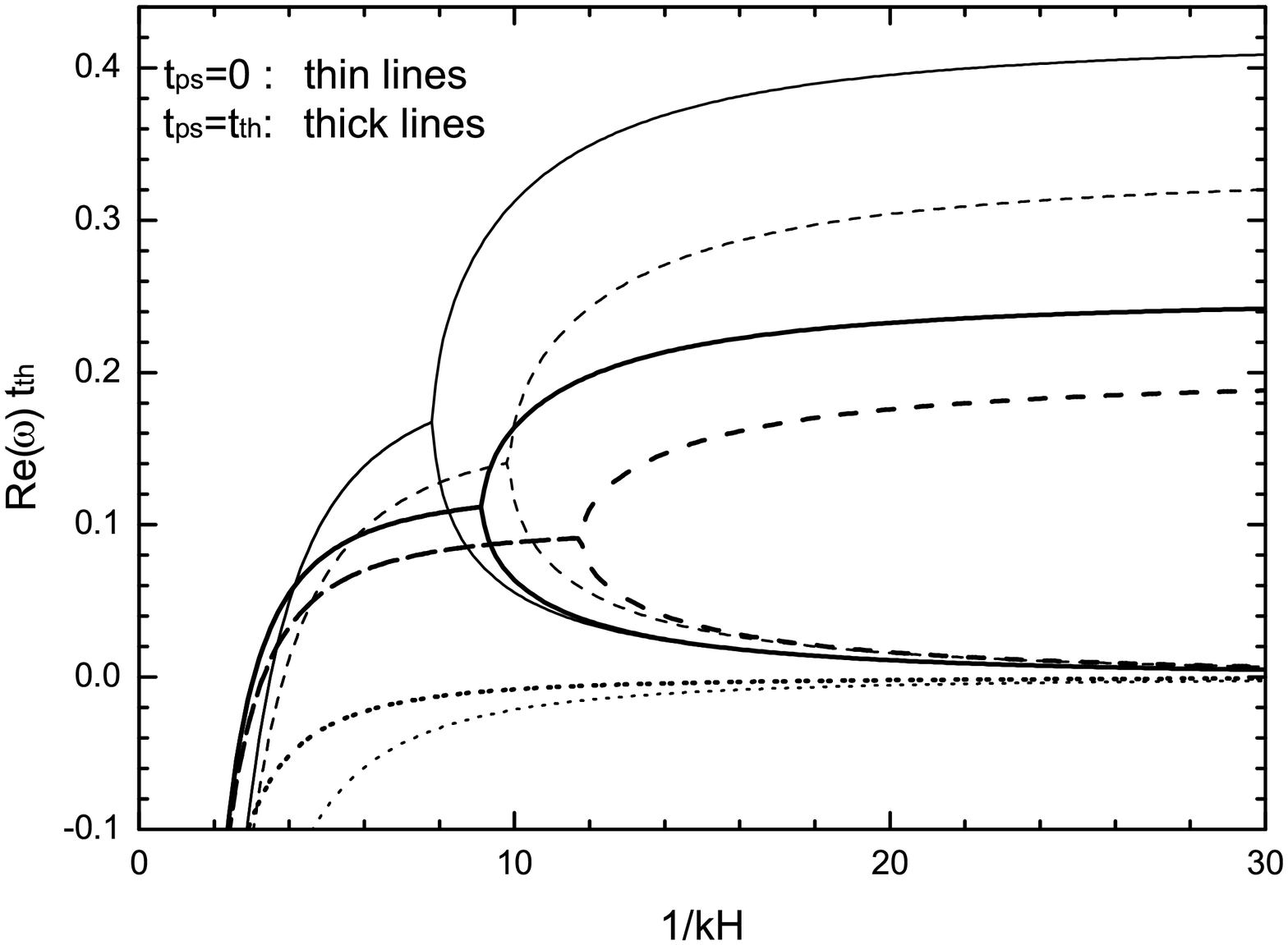}}
\caption{Variation of the growth rate with the radial wavelength
of perturbation.
The thin and thick lines correspond to the standard $\alpha$ stress
prescription and the stress evolution process with $t_{\rm ps} = t_{\rm th}$,
respectively.
The solid, dashed, and dotted lines correspond to $\beta=0.01$, $0.1$,
and $0.4$, respectively.
\label{fig1}}
\end{center}
\end{figure}

\begin{figure}
\begin{center}
{\includegraphics[width=3.00in,height=2.70in]{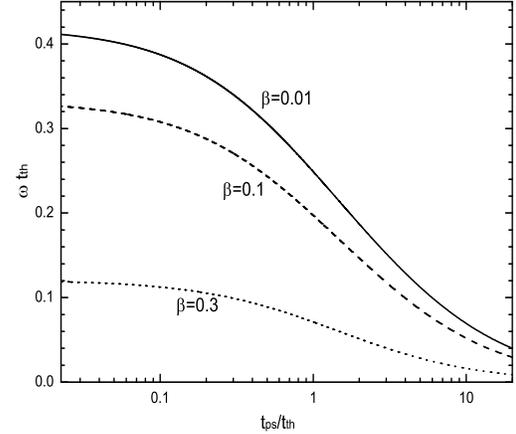}}
\caption{Variation of the growth rate with the timescale of stress delay
in the long-wavelength limit ($kH \to 0$).
\label{fig2}}
\end{center}
\end{figure}

We numerically solve the dispersion equation (\ref{DissipationE})
to obtain the growth rates of unstable modes.
In our calculation, we fix $\alpha = 0.1$ and $\gamma = 5/3$.
There exist three modes for each given radial wavelength of perturbations.
Among the three modes, one is always stable corresponding to a negative
$\omega$, and is therefore neglected in the following analysis.
We study the dispersion relation for two cases: (i) with stress
evolution process ($t_{\rm ps} = t_{\rm th}$); (ii) with the standard
$\alpha$ stress ($t_{\rm ps}=0$).
The growth rates of the remaining two modes are shown in Figure~1.
In this figure, the thick and thin lines correspond to the stress
evolution process and the $\alpha$ stress prescription, respectively.
The solid, dashed, and dotted lines represent the solutions with
$\beta=0.01$, $0.1$, and $0.4$, respectively, where $\beta=0.4$ is
the well known critical value for
the thermal stability with the $\alpha$ prescription.
It is shown that the growth rate of thermally unstable modes (the upper branch)
in the stress evolution scenario is significantly smaller than that in
the $\alpha$ stress case. On the contrary,
the growth rates of the viscously unstable modes
(the lower branch) of the above two cases are similar.
Here, the thermal timescale $t_{\rm th}$ is calculated by
\begin{equation}
t_{\rm th} = \frac{E}{Q_{\rm vis}^+}
= \frac{2[3(1-\beta)(\gamma-1)-\beta]}{3(\gamma-1)}\frac{1}{\alpha\Omega} \ .
\end{equation}
Figure~1 indicates that, even though the stress evolution process does not
change the stability criterion, it may have
significant influence on the growth rate of the thermally unstable mode.

For further investigation, we calculate the solutions under
the long-wavelength limit, i.e., $kH \to 0$. In such case,
the non-zero roots of the dispersion equation (\ref{DissipationE}) are,
\begin{equation}\label{GrowthRate}
\omega = {\frac{1}{2A t_{\rm th}}}
\left [-(\frac{A t_{\rm th}}{t_{\rm ps}}+1)
{\pm} \sqrt{(\frac{At_{\rm th}}{t_{\rm ps}}+1)^2
+ 2\frac{A t_{\rm th}}{t_{\rm ps}}\frac{2-5\beta}{1+\beta}} \right ] \ ,
\end{equation}
where
\begin{displaymath}
A = \frac{\left ( \frac{7-6\beta}{2}+\frac{\beta}{\gamma-1}
- \frac{7}{2}\beta\frac{4-3\gamma}{\gamma-1}\frac{1-\beta}{4-3\beta} \right )}
{[3(1-\beta)+\frac{\beta}{\gamma-1}]}
\frac{4-3\beta}{2(1+\beta)} \ .
\end{displaymath}
The above equation shows the relationship between the growth rate
$\omega$ and the timescale of stress delay $t_{\rm ps}$.

For the case of $t_{\rm ps}/t_{\rm th} \gg 1$, Eq.~(\ref{GrowthRate})
is simplified as
\begin{equation}
\omega = {\frac{1}{2t_{\rm ps}}}
\frac{2-5\beta}{1+\beta} \sim {\frac{1}{t_{\rm ps}}} \ ,
\end{equation}
which means that if $t_{\rm ps}$ is sufficiently larger than $t_{\rm th}$,
the growing timescale of thermal instability is around $t_{\rm ps}$.
On the contrary, for the case of $t_{\rm ps}/t_{\rm th}{\ll}1$,
Eq.~(\ref{GrowthRate}) is reduced to
\begin{equation}
\omega = {\frac{1}{2At_{\rm {th}}}} \frac{2-5\beta}{1+\beta}
\sim {\frac{1}{t_{\rm th}}} \ .
\end{equation}
The above growth rate is exactly the same as that in the $\alpha$ stress case,
in which the growing timescale of thermal instability is around
$t_{\rm th}$.

Figure~2 shows the variation of the growth rate of
the thermally unstable mode in the long-wavelength limit with
$t_{\rm ps}$ for $\beta = 0.01$, 0.1, and 0.3.
The figure clearly illustrates that, for the case of
$t_{\rm ps} \ga t_{\rm th}$,
the stress evolution process will have essential effects on the thermal
instability since the growth rate is significantly decreased. 

\section{Conclusions and discussion}

In this paper, the stress evolution process, which was ignored
in previous works, is taken into account in the linear stability analysis
of standard thin accretion discs.
We show the variation of the growth rate of thermally unstable modes
with the wavelength of perturbations for two cases: with and without stress
delay. We find that the growth rate with stress delay can be
apparently lower than that without stress delay.
We also make some analytical approximation for the growth rate in
long-wavelength limit, and present the relationship between the
specific growth rate and the timescale of stress delay.
In conclusion, the stress evolution process may have essential
influence on the thermal instability by significantly decreasing
the growth rate, in particular for the case in which
the timescale of stress delay is comparable to or even larger than
the thermal timescale.

In a real system the timescale of stress delay remains
unclear. We would argue that this timescale is possibly
comparable to the thermal timescale as follows.
In simulations with an initially weak toroidal or poloidal magnetic field,
the magnetic energy may first experience
an exponential growth during the first few orbits due
to the linear instability, and then will be followed by the nonlinear evolution.
Finally, a saturated quasi-steady state phase may form.
Totally, it will take $15 \sim 20$ orbits for the system to enter
a fully turbulent state (e.g., \citealp{Hawley96,Fromang07}). 
This implies that the evolution timescale of the 
magnetoturbulence is around $15 \sim 20$ times of the dynamical timescale
$t_{\rm dyn}$, which can be regarded as a duration around
the thermal timescale $t_{\rm th}$
(e.g., $t_{\rm th} \sim t_{\rm dyn}/\alpha$ and $\alpha = 0.1$).

As mentioned in the first section, when the theory of the limit cycle behaviour
between the standard thin disc and the slim disc is applied to the
observational quasi-periodic variability of GRS 1915, there exists
some conflict between the theory and the observation
on the duration ratio $t_{\rm high}/t_{\rm low}$ and
the luminosity ratio $L_{\rm high}/L_{\rm low}$.
In our opinion, the stress evolution process may improve the theory
to explain observations
due to the following reasons. When the flow suffers thermal instability,
the growth rate of $\dot M$ can be significantly decreased by
the stress delay.
We therefore can expect a relatively lower $\dot M$
for the outburst phase, thus a corresponding lower $L_{\rm high}/L_{\rm low}$.
As a consequence, for a certain fixed mass accretion rate supply at outer
boundary, if $\dot M$ at outburst phase drops, the duration of this phase
will therefore become longer, thus a moderate value of
$t_{\rm high}/t_{\rm low}$ should appear.
In other words, the influence of the stress delay
on the growth rate may result in moderate ratios of duration and luminosity
of the outburst phase to the quiescent one,
which will be more likely to explain the observational results than
previous calculations. The further investigation
on this issue may require time-dependent numerical calculations including
the stress evolution process.  

Moreover, we would point out the possible application of the stress evolution
process to another geometrically thin disc,
namely the Shapiro-Lightman-Eardley disc (the SLE disc, Shapiro et al.~1976),
which was originally introduced as the inner region of a thin
disc to provide hard X-ray emission. However, it is known that the SLE disc
is viscously stable but suffers thermal instability.
If the stress evolution process is taken into
consideration in the SLE model, and the timescale of stress delay is 
comparable to the thermal timescale, we can therefore expect the existence
of a quasi-stable SLE disc since the decreased growth rate of
thermal instability may not completely destroy the disc and the viscous
stability may help to suppress the thermal instability
in the viscous timescale.

\section*{Acknowledgments}

We thank Shoji Kato and Sheng-Ming Zheng for beneficial discussion.
This work was supported by the National Basic Research Program of China
under grant 2009CB824800, and the National Natural Science Foundation
of China under grants 10833002 and 11073015.

\bsp

\label{lastpage}

\end{document}